\documentstyle[11pt,aaspp4,flushrt]{article} 
\input psfig
\begin{document}
\title{To Probe a Black Hole} 
\author{Andrei Gruzinov}
\affil{Institute for Advanced Study, School of Natural Sciences, Princeton, NJ 08540}

\begin{abstract}
We show how an idealized high-frequency VLBI can probe the metric of nearby supermassive black holes. No quantitative assumptions about plasma dynamics are used in the suggested diagnostic, which is based on strong-field lensing.

\end{abstract}

\section{Introduction}
Sagittarius A* is the largest black hole as seen from Earth, with $2R_S/D=15\mu {\rm as}$, where $R_S$ is the Schwarzschild radius and $D$ is the distance to the Galactic Center (Genzel et al 2000, Ghez et al 1998). Several other nearby supermassive black holes have angular diameters $\sim 10 \mu {\rm as}$  (Richstone et al 1998). 

$15\mu {\rm as}$ is the angular resolution of a 375 GHz VLBI, and the interstellar scattering size towards Sgr A* at 375 GHz is only $\approx 6\mu {\rm as}$. Sgr A* might have already been detected by a 215 GHz VLBI (Krichbaum et al 1998). Radio-astronomers believe that future observations will resolve the black hole (Doeleman \& Krichbaum 1999, Falcke et al 1999). The high-frequency VLBI radio map can show the light capture cross section (the angular diameter is $38\mu {\rm as}$ \footnote{For simplicity, here and in the rest of the paper, we assume that the black hole is not rotating. For concreteness, we discuss Sgr A*.}), the last stable orbit ($45\mu {\rm as}$), the innermost jet formation region, or the combination of the above. 

When the $\sim 40\mu {\rm as}$ structure is detected, one will want to use the radio map to probe the black hole metric. The standard view is that this is impossible. What we see is determined by both the metric and the emissivity. The emissivity depends on the plasma dynamics. The dynamics of a magnetized collisionless plasma in a black hole metric will not be understood in a foreseeable future. Even in Newtonian approximation, our understanding of the low-luminosity accreting black holes is rudimentary (Shvartsman 1971, Gruzinov 1997, 1999, Quataert \& Gruzinov 1999, 2000 (a,b)). 
  
Nevertheless, an idealized high-frequency VLBI can probe the black hole metric. This is possible if the emissivity is variable in space and time. Lensing of this variable emission should be detectable, and can be used to probe the metric.

In \S 2 we discuss one of many possible ways to detect lensing, and therefore to probe the black hole. Only  real data can tell which method would work best. The aim of this paper is to show that high-frequency VLBI can probe  black holes in a plasma-dynamics-independent way -- in principle.

\section{Lensed flares}

The size of Sgr A$^*$ is known to be small. The observed size is scattering-dominated at lower frequencies, up to about 100 GHz. From 215 GHz VLBI, the radius of the emitting region is thought to be in the range 3 -- 13 $R_S$. The flux probably peaks at about 1000 GHz (Zylka et al 1992, Serabyn et al 1997). We will assume that most of the high-frequency radio emission is from a few $R_S$ region around the hole. In what follows we make some further assumptions about the emissivity. These assumptions are only qualitative, the suggested diagnostic does not use any quantitative model of the plasma flow.

Define a radio flare as a sudden brightening of a small region. Here sudden means $\lesssim R_S/c\sim 0.5{\rm min}$ and small means $\lesssim R_S/D\sim 10\mu {\rm as}$. We will assume that VLBI has a comparable angular and time resolution \footnote{Good uv-coverage is a plus but not a must.}.

Every flare is lensed by the hole an infinite number of times, but the flux even in the second image is likely to be very small if the source is far from the hole. The situation is different for sources close to the hole. At $4R_S$ say, the probability that the magnification of the second image is greater than 0.2 is about 10\% (Appendix A). 

If flares do occur, lensed double flares will be detected. If the background emission is detected in the same experiment, one can fix a position (or an assumed position) of the hole in the sky. Given relative positions of the two images and the hole, one can calculate the theoretical time lag (Appendix A). We probe the metric by comparing the theoretical and the observed time lags. For example, a source at $R=5R_S$ and $\alpha =0.2$ (fig. 1) will be seen as a double (magnifications $\approx 2$ and $\approx 1$). The brighter image is 32$\mu$as from the hole, the second image is 28$\mu$as from the hole. The second image appears 50s later.

\acknowledgements This work was supported by the W. M. Keck Foundation and NSF PHY-9513835. I thank Eliot Quataert, John Bahcall, and Avi Loeb for useful discussions.

\begin{appendix}

\section{Lensing in the Schwarzschild metric}

\begin{figure}[htb]
\psfig{figure=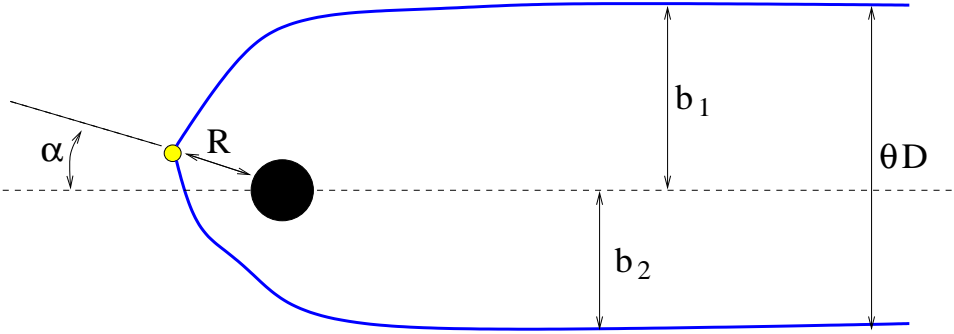,width=5in}
\caption{Lensing of a source at $R$, $\alpha$. $D$ is the distance to the observer, $\theta$ is the angular separation of the double image, $b_1$ is the impact parameter of the brightest image, $b_2$ is the impact parameter of the second brightest image.}
\end{figure}

We use dimensionless units, $R_S=c=1$. In Schwarzschild spherical coordinates $(r,\theta ,\phi )$, the observer is at $(D,0,0)$, where $D\gg 1$. The source is at $(R, \pi -\alpha ,0)$, as shown in figure 1. Only the two brightest images are considered.

Light propagation is given by (eg. Shapiro \& Teukolsky 1983)
\begin{equation}
\dot{r} =b(1-r^{-1})\sqrt{b^{-2}-r^{-2}+r^{-3}} ,
\end{equation}
\begin{equation}
\dot{\psi } =b(1-r^{-1})r^{-2}.
\end{equation}
Where the overdot is the time-at-infinity derivative, $b$ is the impact parameter, and $\psi$ is the polar angle in the plane of motion. The impact parameters of the two brightest images are given by (fig.1.):
\begin{equation}
\pi \pm \alpha = \left( \int _0^{x_2}+\int_{R^{-1}}^{x_2}\right) {dx\over \sqrt{b^{-2}-x^2+x^3} },
\end{equation}
where $x_2$ is the intermediate root of the polynomial $b^{-2}-x^2+x^3$. The minus sign in (A3) corresponds to the brighter image. The propagation time to infinity is
\begin{equation}
t= \left( \int _0^{x_2}+\int_{R^{-1}}^{x_2}\right) {dx\over b(1-x)x^2\sqrt{b^{-2}-x^2+x^3} }.
\end{equation}
The integral diverges, but the time lag $t=t_2-t_1$ is finite. The magnification is given by the ratio of the solid angles of emitted and lensed rays,
\begin{equation}
\mu ={d(\cos \beta )\over d(\cos \alpha )},
\end{equation}
where the emission angle $\beta$, as measured by an observer at rest at $R$, is given by (Shapiro \& Teukolsky 1983)
\begin{equation}
\sin \beta = bR^{-1}\sqrt{1-R^{-1}}.
\end{equation}

\end{appendix}

\end{document}